\documentclass[prl,showpacs,aps,floatfix]{revtex4}

 \usepackage{amsbsy}
\newcommand{\be}{\begin{equation}}
\newcommand{\ee}{\end{equation}}
\newcommand{\brr}{\begin{eqnarray}}
\newcommand{\er}{\end{eqnarray}}

\begin{document}
\title[Physical properties of the Schur complement of local covariance matrices]{Physical properties of the Schur complement of local covariance matrices}

\author{L.F. Haruna and M.C. de Oliveira\footnote{marcos@ifi.unicamp.br}}
\affiliation{Instituto de F\'\i sica ``Gleb Wataghin'', Universidade Estadual de Campinas \\ 13083-970, Campinas, S\~ao Paulo, Brazil.}
\begin{abstract}
General properties of global covariance matrices representing bipartite
Gaussian states can be decomposed into properties of local covariance
matrices and their Schur complements. We demonstrate that given a bipartite Gaussian state $\rho_{12}$
described by a $4\times 4$ covariance matrix \textbf{V}, the Schur
complement of a local covariance submatrix $\textbf{V}_1$ of it can be
interpreted as a new covariance matrix representing a Gaussian operator of party 1 conditioned to local parity measurements on party 2. The connection with a partial parity measurement over a bipartite quantum state and the determination of the reduced Wigner function is given and an operational process of parity measurement is developed. Generalization of this
procedure to a $n$-partite Gaussian state is given and it is demonstrated that the
$n-1$ system state conditioned to a partial parity projection is
given by a covariance matrix such as its $2 \times 2$ block elements are
Schur complements of special local matrices.
\end{abstract}
\maketitle
\section{Introduction}
Several aspects of the Quantum Information theory rely on the ability to prepare and manipulate the quantum state of a given physical system. Obviously this ability is related to the  available physical operations on the actual state encoding. For quantum protocols whose state encoding is spanned by a two dimensional Hilbert space (qubits) one has to be able to completely describe the actual physical operations acting directly on the physical system codifying state space. Thus any physical operation (including measurements) is defined through completely positive maps. For continuous variable encoding as given by Gaussian states\footnote{A Gaussian state is the one which can be fully described by its first and second moments, the latter of which are commonly collected in the covariance matrix.} physical operations given by positive maps can or cannot keep the Gaussian character of the system state. Those operations that map every Gaussian input state into a Gaussian output state are called Gaussian operations \cite{plenio}. For unitary operations acting on a single system Gaussian state, for example, the positive map is completely described by a local $Sp(2,R)$ symplectic (canonical) transformation over the system's covariance matrix. Conversely, as a consequence of the Stone-von Neumann theorem, to every symplectic transformation over the covariance matrix there exists a unique unitary operation acting on the corresponding state space. Those operations correspond to the transformations induced by linear (active and passive) optical devices (beam-splitters, phase-shifters, and squeezers). It is thus not surprising that continuous variable quantum information protocols have been implemented with a Gaussian state encoding and Gaussian operations \cite{braustein,furusawa}.

The whole set of unitary operations does not describe the most general transformation on quantum system states.
A general evolution map of a quantum system including coupling to ancillas (or reservoir) and measurement
is given by quantum operations \cite{kraus}.
 In this formalism the map
\begin{equation}
\rho\rightarrow\frac{\varepsilon(\rho)}{Tr[\varepsilon(\rho)]},
\end{equation}
connects the initial (input) to the final (output) state. The quantum operation $\varepsilon$ is a linear,
trace decreasing map that preserves positivity. A quantum operation $\varepsilon$ satisfying complete positivity
is written as $
\varepsilon(\rho)=\sum_jA_j\rho A_j^\dagger$, 
where ${A_j}$ is a set of system operators, which must satisfy $
\sum_jA_j^\dagger A_j\le I $.
Generally speaking, ``interactions'' with an ancilla (or reservoir) satisfy the trace-preserving condition
$\sum_j A_j^\dagger A_j=I$, while quantum measurements satisfy $\sum_j A_j^\dagger A_j<1 $. Those operations which are trace preserving and irreversible are sometimes named as quantum channels. Gaussian quantum channels \cite{eisertwolf} have the additional feature to keep the Gaussian nature of the system(s) quantum state.


Several authors have described Gaussian quantum channels and operations throughly \cite{plenio,eisertwolf,cirac1,eisert,fiurasek, marcos1,marcos2}.
Regardless the specific interest in each one of those references, one common interesting feature observed is that the Schur
complement \cite{horn} of square matrices representing Gaussian states covariances embodies a manifestation of a physical
operation when considering partial projections and trace operations onto Gaussian states \cite{eisert,fiurasek}.
However the process that delivers the Schur complement of the covariance matrix itself for an input two-mode bipartite Gaussian state has not been discussed yet. As we show in this paper, this is only possible when non-positive operations are allowed. The purpose of the present work is to derive this process for an arbitrary input two-mode bipartite Gaussian state and generalize it for the case of a $n$ partite ($n$-mode)  Gaussian state. Particularly we show that the Schur complement of one of the local covariance matrices of a bipartite Gaussian state appears as the covariance matrix describing a Gaussian operator, which embodies a parity measurement over the other mode of the input state.

We begin by reviewing in Sec. 2 some properties of bipartite Gaussian states \cite{englert} and we describe the decomposition of positivity for a covariance matrix through the Schur complement \cite{horn} structure. In Sec. 3 we discuss examples of Gaussian quantum operations, consisting of the vacuum state projection on one mode of a two-mode bipartite input state, and the partial trace operation. In Sec. 4 we develop our central result, namely the covariance matrix that represents the Schur complement of one of the input subsystem covariance matrix. In Sec. 5 we describe the parity measurement process and the physical properties behind the mathematical quantity as given in Sec. 4. In Sec. 6 we outline how such a Schur complement of a local covariance matrix can be employed for entanglement and global purity quantification. In Sec. 7 we provide a generalization of this local parity measurement for a Gaussian state with $n$-modes in terms of an output covariance matrix. Finally Sec. 8 concludes the paper.

\section{Two-Mode Bipartite Gaussian States}
Any two-mode bipartite quantum state, $\rho_{12}$, is Gaussian if we can write it as\footnote{Throughout this paper we consider the first order moments set to zero with no consequence to the generality of the results, since this can always be performed  with local displacements without altering the state entanglement properties.}
\be
\rho_{12}=\int d\textbf{z} \ e^{\textbf{z}^\dagger\textbf{E}\textbf{a}}e^{-\frac{1}{2}\textbf{z}^\dagger\textbf{V}\textbf{z}}. \label{def_gauss}
\ee
Alternatively, if its symmetric characteristic function is given by $\chi(\textbf{z})=Tr[D(\textbf{z})\rho]=e^{-\frac{1}{2}\textbf{z}^\dagger\textbf{V}\textbf{z}}$, where $D(\textbf{z})=e^{-\textbf{z}^\dagger\textbf{E}\textbf{a}}$ is a displacement operator with $\textbf{z}^\dagger=(z_1^*,z_1,z_2^*,z_1)$, $\textbf{a}^\dagger=(a_1^\dagger,a_1,a_2^\dagger,a_1)$, being $a_1 (a_1^\dagger)$ and $a_2 (a_2^\dagger)$ the annihilation (creation) operators for modes 1 and 2, respectively. Remark that here, as in many integrals through this paper the following notation is employed: $d\textbf{z}=d\textbf{z}_1d\textbf{z}_2$, where $d\textbf{z}_i=d^2z_i=dIm(z_i)dRe(z_i)$ for $i=1,2$. Also,
\begin{equation}
\mathbf{E}=\left( \begin{array}{cc}
\textbf{Z} & \textbf{0} \\
\textbf{0} & \textbf{Z} \end{array} \right), \ 
\textbf{Z}=\left( \begin{array}{cc}
1 & 0 \\
0 & -1 \end{array} \right),
\end{equation}
and \textbf{V} is the Hermitian $4\times 4$ covariance matrix describing all the second order moments $V_{ij}=(-1)^{i+j}\langle\lbrace v_i,v_j^\dagger\rbrace\rangle /2$, where $(v_1,v_2,v_3,v_4)=(a_1,a_1^\dagger,a_2,a_2^\dagger)$, given by
\begin{equation}
\textbf{V}=\left( \begin{array}{cc}
\textbf{V}_1 & \textbf{C} \\
\textbf{C}^\dagger & \textbf{V}_2 \end{array} \right)=\left( \begin{array}{cccc}
n_1+\frac{1}{2} & m_1 & m_s & m_c \\
m_1^* & n_1+\frac{1}{2} & m_c^* & m_s^* \\
m_s^* & m_c & n_2+\frac{1}{2} & m_2 \\
m_c^* & m_s & m_2^* & n_2+\frac{1}{2} \\ \end{array} \right), \label{covariancia_2d}
\end{equation}
where $\textbf{V}_1$ and $\textbf{V}_2$ are $2\times 2$ Hermitian matrices containing only local elements while \textbf{C} is the correlation between the two parties. Furthermore any covariance matrix must be positive semidefinite ($\textbf{V}\geq \mathbf{0}$) and for the operator (\ref{def_gauss}) to represent a physical state the inequality
\be
\textbf{V}+\frac{1}{2}\textbf{E}\geq \mathbf{0}, \label{physical}
\ee
must hold, which is nothing but the fundamental uncertainty principle.

Those general positivity criteria can all be decomposed into properties of the block matrices $\textbf{V}_1, \textbf{V}_2$ and $\textbf{C}$. A convenient decomposition procedure is given by the Schur complement for non-singular block matrices\footnote{Local covariance matrices, $\textbf{V}_1$ and $\textbf{V}_2$ are always non-singular due to the presence of the vacuum fluctuation term $1/2$ in all their diagonal elements.}. In this way for the covariance matrix (\ref{covariancia_2d}), $\textbf{V}\geq \mathbf{0}$ if and only if
\be
\textbf{V}_2\geq \mathbf{0}
\ee
and the Schur complement of $\textbf{V}_2$, given by
\be
S(\textbf{V}_2)=\textbf{V}_1-\textbf{C}\textbf{V}_2^{-1}\textbf{C}^\dagger\geq \mathbf{0}.
\ee
Similar properties have to be satisfied for the generalized uncertainty principle \cite{marcos1}:\be \label{xcond}{\bf
V_2}+\frac 12 {\bf Z}\ge 0,\ee and
 \be \label{xcond2}\left({\bf V_1+\frac 12 {\bf Z}}\right)-{\bf C}\left({\bf
V_2}+\frac 12 {\bf Z}\right)^{-1}{\bf C}^\dagger\ge 0.\ee The emergence of the Schur complement structure and its importance can be fully appreciated through the well known determinant decomposition property. Given $\mathbf{V}$ as described above,
\brr\det{\textbf{V}}&=&\det{\textbf{V}_1} \det({\textbf{V}_2-\textbf{C}^\dagger\textbf{V}_1^{-1}\textbf{C}})\nonumber\\
&=&\det{\textbf{V}_2} \det({\textbf{V}_1-\textbf{C}\textbf{V}_2^{-1}\textbf{C}^\dagger}).\er
Thus, it is immediate that any operation that yields $\textbf{V}_1$ and $S(\textbf{V}_1)$ (or $\textbf{V}_2$ and $S(\textbf{V}_2)$ by symmetry) can be employed, in priciple, to quantify some physical properties of the global covariance matrix $\textbf{V}$. This is the case for entanglement and global purity of a bipartite system, as discussed in details in Ref. \cite{protocolo}, whose central results are commented latter.

\section{Gaussian Operations}

Let us remark on two examples of Gaussian operations, which were previously considered in details in Ref. \cite{plenio,eisert,fiurasek} to show that Schur
complements of block matrices representing Gaussian states
covariances embodies a manifestation of a physical
operation when considering partial projections onto Gaussian
states.

The first example considered is a vacuum state projection over the mode 2 of a two-mode Gaussian state $\rho_{12}$:
\be
\sigma_1^{(0)}=Tr_2\{\vert 0\rangle_2\langle 0\vert\rho_{12}\}. \label{proj_vacuo}
\ee

Writing the Gaussian state form of the one mode vacuum state according to equation (\ref{def_gauss}) in (\ref{proj_vacuo}), we have

\be
\sigma_1^{(0)}=Tr_2\{d\textbf{z}_2 \ e^{-\frac{1}{4}\textbf{z}_2^\dagger\textbf{z}_2}  e^{-\textbf{z}_2^\dagger\textbf{Z}\textbf{a}_2}\rho_{12}\}.
\ee

Now replacing the Gaussian form for $\rho_{12}$ (\ref{def_gauss}) and evaluating the trace, we have that
\begin{eqnarray}
\sigma_1^{(0)} &=& \int d\textbf{z}_1 \ e^{\textbf{z}_1^\dagger\textbf{Z}\textbf{a}_1}\int d\textbf{z}_2 \ e^{-\frac{1}{2}\textbf{z}^\dagger\textbf{V}^{(0)}\textbf{z}}\nonumber \\
&=&
\frac{1}{\sqrt{\det\left(\textbf{V}_2+\frac{1}{2}\textbf{I}\right)}}\int d\textbf{z}_1 \ e^{\textbf{z}_1^\dagger\textbf{Z}\textbf{a}_1} e^{-\frac{1}{2}\textbf{z}_1^\dagger\left[\textbf{V}_1-\textbf{C}\left(\textbf{V}_2+\frac{1}{2}\textbf{I}\right)^{-1}\textbf{C}^\dagger\right]\textbf{z}_1},
\end{eqnarray}
with $d\textbf{z}_1=d^2z_1=dIm(z_1)dRe(z_1)$. We notice that the resulting state has also a Gaussian form with covariance matrix given by (\ref{schur_vacuo}) 
\be
{\bf\Gamma}_1^{(0)}=\textbf{V}_1-\textbf{C}\left(\textbf{V}_2+\frac{1}{2}\textbf{I}\right)^{-1}\textbf{C}^\dagger. \label{schur_vacuo}
\ee
Here \textbf{I} is a $2\times 2$ identity matrix.
The covariance matrix of the reduced state, (\ref{schur_vacuo}) is the Schur complement of the matrix
\be
\textbf{M}_1=\left( \begin{array}{cc}
\textbf{V}_1 & \textbf{C} \\
\textbf{C}^\dagger & \textbf{V}_2+\frac{1}{2}\textbf{I} \end{array} \right)
\ee
in relation to $\textbf{V}_2+\frac{1}{2}\textbf{I}$ \cite{plenio}.

In the second example we consider a simple partial trace operation:
\be
\rho_1=Tr_2\{\rho_{12}\}. \label{trace}
\ee In this case it is obvious that $\rho_1$ is still a Gaussian state and that it will be given, in terms of the representation \ref{def_gauss}, as
\be \rho_1=\int d\textbf{z}_1 e^{\textbf{z}_1^\dagger\textbf{Z}\textbf{a}_1}e^{-\frac{1}{2}\textbf{z}_1^\dagger\textbf{V}_1\textbf{z}_1}.
\ee
Now, the covariance matrix of the reduced state is given  by the Schur complement of  matrix $\textbf{V}_2$ as given by the matrix
\be
\textbf{M}_2=\left( \begin{array}{cc}
\textbf{V}_1 & \textbf{0} \\
\textbf{0} & \textbf{V}_2 \end{array} \right).
\ee

\section{Emergence of the Schur Complement Structure of Local Covariance Matrices}

As we noted in the previous section, projection operations followed by partial trace operations are manifested as Schur complements of  $4\times 4$ matrices. However none of those matrices is the covariance matrix of the input bipartite Gaussian state since the vacuum state covariance matrix must be inserted, or the correlation matrix $\textbf{C}$ neglected. Therefore, we need a physical measurement process that is able to yield the exact Schur complement of one of the local covariance matrices. In this section we describe the mathematical operation which is able to realize this task and in the next section we describe the physical operation behind it.

\vspace*{12pt}
\noindent
{\bf Proposition~1:} Given a bipartite Gaussian state $\rho_{12}$, the covariance matrix $\Gamma_1$ describing the Gaussian operator $\sigma_1$ of mode 1 conditioned to a parity projection over the mode 2,
\be
\sigma_1 = Tr_2\left\{e^{i\pi a_2^\dagger a_2}\rho_{12}\right\}, \label{est_sigma1}
\ee
is given by the Schur complement of the covariance matrix \textbf{V} of the input bipartite state $\rho_{12}$ in relation to $\textbf{V}_2$:
\be
{\bf\Gamma}_1=\textbf{V}_1-\textbf{C}\textbf{V}_2^{-1}\textbf{C}^\dagger. \label{schur_rel1}
\ee

\vspace*{12pt}
\noindent
{\bf Proof:}
We can write the parity operator as an integral over the displacement operator \cite{davidovich,wodkiewicz}, in this case the equation (\ref{est_sigma1}) can be rewritten as
\be
\sigma_1 = \frac12Tr_2\left\{\int \ d\textbf{s}_2 e^{-\textbf{s}_2^\dagger\textbf{Z}\textbf{a}_2}\rho_{12}\right\} \label{est_schur},
\ee with $d\textbf{s}_2=d^2s_2=dIm(s_2)dRe(s_2)$.
As $\rho_{12}$ represents a Gaussian state we can write it in the form (\ref{def_gauss}) and replacing this Gaussian form in (\ref{est_schur}) we have
\begin{eqnarray}
\sigma_1 &=& \frac12 Tr_2\left\{\int d\textbf{s}_2 e^{-\textbf{s}_2^\dagger\textbf{Z}\textbf{a}_2}\int d\textbf{z}e^{\textbf{z}^\dagger\textbf{E}\textbf{a}}e^{-\frac{1}{2}\textbf{z}^\dagger\textbf{V}\textbf{z}}\right\} \nonumber \\
&=&\frac12
Tr_2\left\{\int d\textbf{s}_2 e^{-\textbf{s}_2^\dagger\textbf{Z}\textbf{a}_2}\int d\textbf{z}_1\int d\textbf{z}_2 e^{\textbf{z}_1^\dagger\textbf{Z}\textbf{a}_1}e^{\textbf{z}_2^\dagger\textbf{Z}\textbf{a}_2}e^{-\frac{1}{2}\textbf{z}^\dagger\textbf{V}\textbf{z}}\right\} \nonumber \\
&=&\frac12
\int d\textbf{s}_2\int d\textbf{z}_1\int d\textbf{z}_2 Tr_2\left\{e^{(\textbf{z}_2-\textbf{s}_2)^\dagger\textbf{Z}\textbf{a}_2}\right\}  e^{-\frac{1}{2}\textbf{z}_2^\dagger\textbf{Z}\textbf{z}_2}
\nonumber \\
& &\times
e^{\textbf{z}_1^\dagger\textbf{Z}\textbf{a}_1}e^{-\frac{1}{2}\textbf{z}^\dagger\textbf{V}\textbf{z}} e^{\frac12\textbf{z}_2^\dagger\textbf{Z}\textbf{s}_2},
\end{eqnarray}
but
\be
Tr_2\left\{e^{(\textbf{z}_2-\textbf{s}_2)^\dagger\textbf{Z}\textbf{a}_2}\right\}=e^{-\frac{1}{2}\vert z_2-s_2\vert^2}\delta^{(2)}(\textbf{z}_2-\textbf{s}_2),
\ee
therefore
\begin{eqnarray}
\sigma_1 
&=&\frac12
\int d\textbf{z}_1 e^{\textbf{z}_1^\dagger\textbf{Z}\textbf{a}_1}\int d\textbf{z}_2e^{-\frac{1}{2}\textbf{z}^\dagger\textbf{V}\textbf{z}}\nonumber \\
&=&\frac12
\frac{1}{\sqrt{\det{\textbf{V}_2}}}\int d\textbf{z}_1 e^{\textbf{z}_1^\dagger\textbf{Z}\textbf{a}_1}e^{-\frac{1}{2}\textbf{z}_1^\dagger\left(\textbf{V}_1-\textbf{C}\textbf{V}_2^{-1}\textbf{C}^\dagger\right)\textbf{z}_1},
\end{eqnarray}
where we notice that the resulting state has also a Gaussian form with covariance matrix given by (\ref{schur_rel1}).
\vspace*{12pt}
\noindent

\section{Expected Parity and Probability Distribution Functions}

In the previous section we have given a mathematical connection of the Schur complement of a local covariance matrix $\textbf{V}_2$ with the parity operation $e^{i\pi a_2^\dagger a_2}$ over mode 2 of a two-mode Gaussian state and further reduction to mode 1. To completely understand the physical meaning of such an operation we must develop a theory for parity measurement detection.

 Indeed, parity operation properties have received considerable attention recently, in connection to experimentally accessible measures, for description of quasiprobability distribution functions for single and entangled systems and for quantum information and computation proposals \cite{alfredo,raul,alfredo2,mcomilburn,mcomunro, dodonov}. Moreover the average parity of a given quantum state is related with its Wigner function \cite{wigner} at the origin of the phase space and nowadays it can be experimentally determined for radiation fields through photocounting experiments \cite{wodkiewicz,banaszek}, or in microwave cavity quantum electrodynamics experiments \cite{davidovich,bertet}

Bellow we give a description of the parity measurement process.
Given a prior quantum state $\rho$, the post-selected state conditioned to a parity ($p=1,-1$) measurement is given by 
\begin{equation}
\rho_{p} =\frac{\mathcal{N}_p\,\,\rho}{Tr\left\{\mathcal{N}_p\,\,\rho\right\}},
\end{equation}\label{operation}
with probability $P_p=Tr\left\{\mathcal{N}_p\,\,\rho\right\}$, where the super-operator $\mathcal{N}_{p}$, $p=1,-1$ is given by\footnote{The construction of this quantum operation is derived inversely by recalling that the Wigner function at the origin of the phase space is the average of the parity operator. Indeed from that one would obtain $\mathcal{N}_{p}\cdot=\left(\sum_{n\in p}|n\rangle\langle n|\right)\cdot\left(\sum_{n\in p}|n\rangle\langle n|\right) $. The completeness relation for the Fock states followed by the fact that $\mathcal{N}_{p}\cdot$ for all pratical purposes is always followed by the trace operation lead to the diagonal form of Eq. (\ref{operation}) convenient for the operator sum representation.}
\begin{equation}
\mathcal{N}_{p}\cdot=\sum_{n\in p}|n\rangle\langle n|\cdot |n\rangle\langle n|,
\end{equation} 
i.e., it indicates projections over even or odd Fock states $|n\rangle$.
For $p=1$ the sum runs over even natural numbers, while if $p=-1$ it runs over odd natural numbers. The probability for the occurrence of the two events write independently as
\brr
P_1=Tr\left\{\mathcal{N}_1\,\,\rho\right\}=\sum_{n_{even}}\langle n|\rho |n\rangle,\\
P_{-1}=Tr\left\{\mathcal{N}_{-1}\,\,\rho\right\}=\sum_{n_{odd}}\langle n|\rho |n\rangle,\er and it can be evidenced that $P_1+P_{-1}=1$, as it should be.
The average parity, $\bar p$, is then simply given by
\begin{equation}
\bar p=\sum_p p P_p=\sum_{n_{even}}\langle n|\rho |n\rangle-
\sum_{n_{odd}}\langle n|\rho |n\rangle=Tr\left\{e^{i\pi a^\dagger a}\rho\right\}.
\ee
The average parity was recognized independently by Grossmann \cite{grossmann} and by Royer \cite{royer} to be proportional to the Wigner distribution function at the origin of the phase space:
\be
W(\mathbf{0})=2\,\, \bar p, \label{wigner_zero}
\ee
The values of this function in other points of the phase-space can be achieved by performing displacements over the input state \cite{cahill,grossmann,royer}, such that
\be
W({\boldsymbol \alpha})=2\sum_{n=0}^\infty(-1)^n \langle n\vert D({\boldsymbol \alpha})\rho D^\dagger({\boldsymbol \alpha})\vert n\rangle,
\ee
where $D({\boldsymbol \alpha})$ is the displacement operator.

This approach can be extended for bipartite systems as given by the density operator $\rho_{12}$. The joint post-selected state $\rho_{12}^{(p)}$ conditioned to a parity ($p=1,-1$) measurement over the mode 2 is given by 
\begin{equation}
\rho_{12}^{(p)} =\frac{\mathcal{N}_p\,\,\rho_{12}}{Tr\left\{\mathcal{N}_p\,\,\rho_{12}\right\}},
\end{equation}
with probability $P_p=Tr_{12}\left\{\mathcal{N}_p\,\,\rho_{12}\right\}$, and the operation $\mathcal{N}_{p}$, now reads as
\begin{equation}
\mathcal{N}_{p}\cdot= \mathbf{I}_1\otimes\sum_{n\in p}|n\rangle\langle n|_2\cdot |n\rangle\langle n|_2.
\end{equation} Now the probability for the occurrence of the two events write as
\brr
P_1=Tr_2\left\{\mathcal{N}_1\,\,\rho_2\right\}=\sum_{n_{even}}\langle n|\rho_2 |n\rangle_2,\\
P_{-1}=Tr_2\left\{\mathcal{N}_{-1}\,\,\rho_2\right\}=\sum_{n_{odd}}\langle n|\rho_2 |n\rangle_2,
\er 
where $\rho_2=Tr_1\{\rho_{12}\}$.
The average parity of mode 2 is related to its Wigner function as
\be
W_2(\mathbf{0})=2\,\, \bar p=2\,\,  Tr_{12}\left\{e^{i\pi a_2^\dagger a_2}\rho_{12}\right\}, \label{wigner_zero1}
\ee
In view of this, the significance of the operator $\sigma_1$ is clearly identified as the difference between the states of mode 1 conditioned to projections on the even Fock subspace and on the odd Fock subspace of mode 2:
\be \sigma_1= \sum_{n_{even}}\langle n|\rho_{12} |n\rangle_2-
\sum_{n_{odd}}\langle n|\rho_{12} |n\rangle_2,\ee\label{sigma}
i.e., the difference between the states of mode 1 conditioned to even and odd parity measurements on mode 2 out of an ensemble of identicaly prepared $\rho_{12}$ states.

Finally, an interesting relation follows from Eq. (\ref{sigma}). It is immediate to check that the Wigner function of mode 2 at the origin of the phase-space can be inferred by
\be W_2(\mathbf{0})=2\,Tr_1\{\sigma_1\}.\ee
In this case, this result becomes very interesting, for it means that in a bipartite Gaussian state, the achievement of the Wigner function at the origin of the phase-space of subsystem 2, is the trace of the Gaussian operator of subsystem 1 $\sigma_1$, as above, and {\it vice versa}. The resulting covariance matrix representing $\sigma_1$, will be given by (\ref{schur_rel1}).

Also, from Eq. (\ref{wigner_zero1}), we can derive a relationship between the Wigner function at the origin of phase space $W_2(0)$ of one mode Gaussian state with covariance matrix $\textbf{V}_2$ and the Schur complement $S(n_2+\frac12)$ of one diagonal element of $\textbf{V}_{2}$, which in this case is a scalar, being
\be
W_2(0)=\frac{1}{\sqrt{\left(n_2+\frac 12\right) S(n_2+\frac12)}},
\ee
where
\be
S(n_2+\frac12)=n_2+\frac12-\frac{\vert m_2\vert^2}{n_2+\frac12}.
\ee
Moreover, we note that the Wigner function of a mode at the origin of the phase-space, and thus $S(n_2+\frac12)$ is directly related with the $I_2=\left(n_2+\frac12\right)^2-\vert m_2\vert^2$ element of the four invariant set of the $Sp(2,R)\otimes Sp(2,R)$ group \cite{marcos1}:  $I_1=\det\textbf{V}_1$, $I_2=\det\textbf{V}_2$, $I_3=\det\textbf{C}$ and $I_4=Tr[\textbf{V}_1\textbf{Z}\textbf{C}\textbf{Z}\textbf{V}_2\textbf{Z}\textbf{C}^\dagger\textbf{Z}]$.

Another way to write $\sigma_1$ is by expanding it in a coherent states basis,
\be
\rho_{12}=\int d^2\beta_1 d^2\beta_2 P(\beta_1, \beta_2)\vert\beta_1,\beta_2\rangle\langle\beta_1,\beta_2\vert \label{rho_P}
\ee
where $P(\beta_1, \beta_2)$ is the Glauber $P$ function of the joint system $\rho_{12}$.
Replacing (\ref{rho_P}) in (\ref{est_sigma1}) and using (\ref{wigner_zero1}) we have
\be
\sigma_1=\int d^2\beta_1\vert\beta_1\rangle\langle\beta_1\vert\int d^2\beta_2 W_c(0)P(\beta_1,\beta_2),
\ee
where $W_c(0)=e^{-2\vert\beta_2\vert^2}$, is the Wigner function of the coherent state $|\beta_2\rangle$ (that spans the base of mode 2) at the origin of phase space. Thus by defining
\be
\mathcal{P}(\beta_1)\equiv\int d^2\beta_2 W_c(0)P(\beta_1,\beta_2)\ee we have
\be
\sigma_1=\int d^2\beta_1\vert\beta_1\rangle\langle\beta_1\vert \mathcal{P}(\beta_1).
\ee
When there is only a reduction of the system without association to a measurement in one subsystem, the $P$ function of the reduced state can be obtained by integrating the bipartite state $P$ function, $P(\beta_1,\beta_2)$, over the traced mode variable. But in (\ref{rho_P}), the $P$ function is associated with a weight in the form of a Gaussian function, where it shows that the values of the $P$ function of $\sigma_1$ over the variables nearly to the origin of the phase space are more important.
The influence in mode 1 by a parity measurement in mode 2 can also be noted if we compare the covariance matrix (\ref{schur_rel1}) with the respective matrix $\textbf{V}_1$ of the reduced operator $\rho_1=Tr_2\{\rho_{12}\}$, indicating that the parity measurement insert global properties in the local terms of mode 1.

\section{Entanglement characterization of a two-mode bipartite Gaussian state}

As we saw in the previous sections, the achievement of the Schur complement structure of a two-mode Gaussian state covariance matrix is related with local parity measurements and the Wigner function at the origin of the phase-space. However, this result has a main physical interpretation related to the entanglement characterization of a two-mode Gaussian state via a protocol based on local operations and classical communication (LOCC) as discussed in Ref. \cite{protocolo}.

The protocol consists basically in the attainment of all local sympletic invariants ($I_1, I_2, \vert I_3\vert$ and $I_4$), discussed in the previous section, via only LOCC and through Proposition 1 of this paper.

Suppose that Alice and Bob share many copies of a two-mode Gaussian state, where the subsystens related to mode 1 are given to Alice and those related to mode 2 to Bob. The quantities $I_1$ and $I_2$ can be locally determined by the reconstruction of the submatrices $\textbf{V}_1$ and $\textbf{V}_2$ or by purity measurements of the mode 1 and 2, performed by Alice and Bob, respectively. If Alice is the one who want to achieve the entanglement characterization, Bob has to send his matrix or purity outcome to Alice via a classical communication channel. $\vert I_3\vert$ and $I_4$ are determined using the Proposition 1, where essencially Alice has to construct the covariance matrix ${\bf\Gamma}_1$ (\ref{schur_rel1}) from the diference between the correlation matrices of the mode 1 subsystem conditioned to projections onto odd and even states of the mode 2 subsystem. For that, Bob has to perform a parity measurement on each subset component letting Alice to know to which copy does that operation corresponds and the respective outcome, i.e., even parity (eigenvalue 1) or odd parity (eigenvalue -1). Alice then separates her copies in two groups, the even (\textit{e}) and the odd (\textit{o}) ones, each one containing the copies conditioned by an even and odd parity measurement on Bob's copies respectively, allowing her to obtain the correlation matrices $\textbf{V}_{1e}$ and $\textbf{V}_{1o}$ related to each group. Consequently, the matrix ${\bf\Gamma}_1$ is obtainned by subtracting the odd correlation matrix from the even one ${\bf\Gamma}_1=\textbf{V}_{1e}-\textbf{V}_{1o}$. With this three matrices ($\textbf{V}_1, \textbf{V}_2$ and ${\bf\Gamma}_1$) in hand, Alice is able to completely characterize the Gaussian state's entanglement content as well as its purity without any global or nonlocal measurements. Two invariants are obtained by $I_1=\det\textbf{V}_1$ and $I_2=\det\textbf{V}_2$, while the third one $\vert I_3\vert$ is calculated from the expression

\be
\vert I_3\vert=\sqrt{I_2\det(\textbf{V}_1-{\bf\Gamma}_1)},
\ee
and the fourth quantity $I_4$ by

\be
I_4=I_1I_2+I_3^2-I_2\det{\bf\Gamma}_1.
\ee

This four quantities allows, for example, decide whether or not the two-mode Gaussian states is entangled. For that, the Simon separability \cite{simon} is the relation used to perform this test, i.e., the state is not entangled if, and only if, 

\be
I_1I_2+\left(\frac{1}{4}-\vert I_3\vert^2\right)^2-I_4\geq(I_1+I_2)/4.
\ee
Moreover, for a symmetric state ($I_1=I_2$), the entanglement can be quantified via the entanglement of formation ($E_f$) \cite{rigolin,giedke}:

\be
E_f(\rho_{12})=f\left(2\sqrt{I_1+\vert I_3\vert-\sqrt{I_4+2I_1\vert I_3\vert}}\right),
\ee
where $f(x)=c_+(x)\log_2[c_+(x)]-c_-(x)\log_2[c_-(x)]$ and $c_\pm(x)=(x^{1/2}\pm x^{1/2})^2/4$. For arbitrary two-mode Gaussian states ($I_1\neq I_2$) we can work with lower bound for $E_f$ \cite{rigolin} or calculate its negativity or logarithmic negativity \cite{vidal}.

Note that all the procedures adopted to achieve the four invariants were made without any type of global measurements or operations, allowing to characterize the two-mode Gaussian state's entanglement content via only local measurements and a classical communication channel.

\section{Parity Measurement in a $N$-Mode Gaussian State}
Now we can generalize the result of theorem 1 for the case of a $n$-mode Gaussian state. 

\vspace*{12pt}
\noindent
{\bf Proposition 2:} Parity measurement in the $m$ mode of a Gaussian state of $n$-modes with covariance matrix $2n\times 2n$ given by
\be
\textbf{V}_{2n\times 2n}=\left( \begin{array}{ccccc}
\textbf{V}_{11} & \textbf{C}_{12} & \textbf{C}_{13} & \ldots & \textbf{C}_{1n}\\
\textbf{C}^\dagger_{12} & \textbf{V}_{22} & \textbf{C}_{23} & & \vdots\\
\textbf{C}^\dagger_{13} & \textbf{C}^\dagger_{23} & \textbf{V}_{33} & & \\
\vdots & & & \ddots & \textbf{C}_{(n-1)n}\\
\textbf{C}_{1n}^\dagger  & \ldots & & \textbf{C}_{(n-1)n}^\dagger & \textbf{V}_{nn}\end{array} \right),
\ee
associate the state  of $n-1$ resulting modes, in such a manner that the resulting covariance matrix $2(n-1)\times2(n-1)$ is formed by $2\times 2$ block matrices, located in line $i$ and column $j$, given by
\be
{\bf\Gamma}_{ij}=\textbf{M}_{ij}-\textbf{M}_{im}\textbf{M}_{mm}^{-1}\textbf{M}_{im}^\dagger, \label{comp_schur_n}
\ee
such that for $i=j$, $\textbf{M}_{ii}=\textbf{V}_i$ is the covariance matrix of the reduced operator for the subsystem $i$, and for $i\neq j$ , $\textbf{M}_{ij}=\textbf{C}_{ij}$ are the matrices that represent the correlations between the $n$ modes of the global system, noting that in this case $\textbf{M}_{ji}=\textbf{M}_{ij}^\dagger$.

\vspace*{12pt}
\noindent
{\bf Proof:}
This proof is made by induction. As we have already derived how a parity measurement in one mode of a bipartite Gaussian state affect the other mode in terms of the covariance matrix (\ref{schur_rel1}), it is possible to derive a simmilar relation for states with 3 and 4 modes and we can verify that there is a standard form between the influence of a parity measurement with the resulting reduced covariance matrices, allowing to make a generalization in the case of a $n$-mode state. 
For a tripartite Gaussian state, $\rho_{123}$, with covariance matrix given by
\be
\textbf{V}_{123}=\left( \begin{array}{ccc}
\textbf{V}_1 & \textbf{C}_{12} & \textbf{C}_{13} \\
\textbf{C}^\dagger_{12} & \textbf{V}_2 & \textbf{C}_{23} \\
\textbf{C}^\dagger_{13} & \textbf{C}^\dagger_{23} & \textbf{V}_3 \end{array} \right),
\ee
the resulting bipartite Gaussian state will be conditioned to a parity measurement in the mode 3 of the global system, $\sigma_{12}=Tr\{e^{i\pi a_3^\dagger a_3}\rho_{123}\}$, such that the respective covariance matrix is 
\be
{\bf\Gamma}_{12}=\left( \begin{array}{cc}
\textbf{V}_1-\textbf{C}_{13}\textbf{V}_3^{-1}\textbf{C}_{13}^\dagger & \textbf{C}_{12}-\textbf{C}_{13}\textbf{V}_3^{-1}\textbf{C}_{23}^\dagger \\
\textbf{C}_{12}^\dagger-\textbf{C}_{23}\textbf{V}_3^{-1}\textbf{C}_{13}^\dagger & \textbf{V}_2-\textbf{C}_{23}\textbf{V}_3^{-1}\textbf{C}_{23}^\dagger \end{array} \right).  \label{comp_schur_3}
\ee
Note that each block element of the matrix above are a Schur decomposition of an another matrix $4\times 4$ and can be obtained by the relation (\ref{comp_schur_n}). 

With an analogous calculus for the case of a 4-mode state, the reduced covariance matrix conditioned to a parity measurement in mode 4 is given by
\be
{\bf\Gamma}_{123}=\left( \begin{array}{ccc}
\textbf{V}_1-\textbf{C}_{14}\textbf{V}_4^{-1}\textbf{C}_{14}^\dagger & \textbf{C}_{12}-\textbf{C}_{14}\textbf{V}_4^{-1}\textbf{C}_{24}^\dagger & \textbf{C}_{13}-\textbf{C}_{14}\textbf{V}_4^{-1}\textbf{C}_{34}^\dagger\\
\textbf{C}_{12}^\dagger-\textbf{C}_{24}\textbf{V}_4^{-1}\textbf{C}_{14}^\dagger & \textbf{V}_2-\textbf{C}_{24}\textbf{V}_4^{-1}\textbf{C}_{24}^\dagger & \textbf{C}_{23}-\textbf{C}_{24}\textbf{V}_4^{-1}\textbf{C}_{34}^\dagger\\
\textbf{C}_{13}^\dagger-\textbf{C}_{34}\textbf{V}_4^{-1}\textbf{C}_{14}^\dagger & \textbf{C}_{23}-\textbf{C}_{34}\textbf{V}_4^{-1}\textbf{C}_{24}^\dagger & \textbf{V}_3-\textbf{C}_{34}\textbf{V}_4^{-1}\textbf{C}_{34}^\dagger 
\end{array} \right). \label{comp_schur_4}
\ee
In the same way of the case of a measurement over the states with 2 and 3 modes, we noted that the block elements of the matrix given by (\ref{comp_schur_4}) can also be described by (\ref{comp_schur_n}).

\section{Conclusion}
In this work, we have investigated which operation over a two-mode bipartite Gaussian state delivers the Schur complement form of an input local covariance matrix. We discovered that parity measurements in one mode of a global bipartite two-mode state influences the other mode in such a manner that there exists a Gaussian operator whose covariance matrix has the form of a Schur complement of one of the local covariance matrices of the input state. This operator is given as the difference between the reduced state of one subsystem conditioned to even and odd projections on the other subsystem. As the parity measurement and the Wigner function are strictly related, it is possible to associate the Schur complement of a local covariance matrix with the process to achieve the Wigner function of one mode and, due to the invariance of the Gaussian properties by displacements, with the achievement of only one point of this function. At the origin, the Wigner function of an one mode Gaussian state is related with one element of the four invariant set of the $Sp(2,R)\otimes Sp(2,R)$ group and so is the Schur complement of a local covariance matrix. Moreover we have generalized the approach for a $n$-partite Gaussian state verifying that after a parity measurement in one mode, the $n-1$ system state has a covariance matrix with $2\times2$ block elements in a form of Schur complements of special block matrices. We believe that our findings have both conceptual and practical implications for the development of continuous variable protocols with Gaussian states \cite{protocolo}.

We are pleased to thank G. Rigolin for valuable discussions. This work is partially supported by FAPESP and by CNPq.
 


\end{document}